# Effective Trade Execution[1]

**RICCARDO CESARI**
Department of Statistics, University of Bologna

**MASSIMILIANO MARZO**
Department of Economics, University of Bologna

**PAOLO ZAGAGLIA**
Department of Economics, University of Bologna

**This version: June 2012**

**ABSTRACT**

This paper examines the role of algorithmic trading in modern financial markets. Additionally, order types, characteristics, and special features of algorithmic trading are described under the lens provided by the large development of high frequency trading technology. Special order types are examined together with an intuitive description of the implied dynamics of the order book conditional to special orders (iceberg and hidden). The chapter provides an analysis of the transaction costs associated with trading activity and examines the most common trading strategy employed in the market. It also examines optimal execution strategy with the description of the Efficient Trading Frontier. These concepts represent the tools needed to understand the most recent innovations in financial markets and the most recent advances in microstructures research.

---

[1] This paper is forthcoming as Chapter 19 in "Portfolio Theory and Management", ed. By Gregg Filbeck and Kent Baker, Oxford University Press, December 2012. We thank Kent Baker and Gregg Filbeck for precious comments. The usual disclaimer applies.

# 1. INTRODUCTION

The traditional view about financial market organizations is that market participants meet to exchange securities in their respective interests and positions. According to this wisdom, intermediaries (brokers or dealers) conduct floor trades with the goal of finding the various matches between market participants. Technological advances represent a critical component relative to the evolution of market organization. The strong development of technology adoption results in an increased speed of financial transactions by gradually reducing the importance of physical location on the trading floor.

Moreover, the search for liquidity, which is the tension of the trading industry towards the minimization of execution costs and trade impact, creates room for an increase in the number of competing trading venues. The resulting international trading landscape is characterized by a substantial amount of automated trading technologies employed in the trading activity and an increase of market fragmentation with the emergence of several trading venues and the consequent emergence of competition for the order flow. The adoption of Algorithmic Trading (AT) and High Frequency Trading (HFT) is the direct consequence of this competition.

This paper explores the role of innovation in trading activity and its impact on the various order characteristics, trading strategies and costs in the recent evolution of market microstructure literature. The rapidly evolving scenario witnessed by the role of trading algorithms results in a very important set of changes in the market architecture, regarding optimal order size, market timing strategies, information disclosure, and competition among trading venues and market participants. Existing microstructure theories are being re-evaluated as a result of the expanding role of AT and HFT.

AT results in implementing a set of trading strategies or "algorithms" that often involve computer-based implementation. An *algorithm* is a set of decision rules and strategies used to satisfy a specific goal. HFT is an evolution of AT. HFT represents the implementation of proprietary trading strategies by agents by adopting fast technological computing to realize trading at very high frequencies and extremely low latencies (below 10 microseconds). The main drivers of AT and HFT can be identified according to the following elements: (1) the market access model; (2) the fee structure; and (3) latency reduction and increases in competition and fragmentation of the order flow.

The market access model has evolved in an important way. In the past, designated intermediaries (brokers or market makers) had the role of collecting orders from their customers and allocated them to appropriate market venues. This business model has evolved over time by qualifying the market access in two ways: (1) direct market access (DMA) and (2) sponsored access (SA). With DMA, investors can directly place their orders in the marketplace by using the broker's infrastructure of access to the market. In this case, brokers do not intervene in the order placement mechanisms: their role is only to provide the infrastructure to their clients to gain access to the market, after only a pre-trade risk check.

With SA, an agent (which could be an investment firm, a private wealthy client, or a mutual fund) can have access to the market without being a member of the floor by using the broker's member ID, without necessarily using the latter's infrastructure. If the trading venue provides filtered SA, then the broker can implement pre-trade risk checks. Conversely, if the market provides "naked' access," then the sponsor does a preliminary check to protect his own risk exposure. The main advantage of SA over DMA is the strong reduction of latency, which makes SA more attractive for AT of HFT operators.

The recent evolution of policies implemented at the level of each market exchange results in a strong increase in fee competition among trading venues and attracts the highest possible quantity of the order flow. Market participants reducing liquidity from the market are charged a higher fee, while traders increasing liquidity are charged a lower fee or are provided with a rebate. BATS, Chi-X and Turquoise, which are some specific trading venues used to raise the competition for the order flow, implemented this fee structure for the first time, Given the implementation of such aggressive trading schemes, all European exchanges have lowered their fees to avoid losing order flow. If an order is incorrectly implemented, substantial risk exists that the order is no longer appropriate in terms of price limit or size, because the order may be executed at an improper price or even not executed at all. In this context, the design of a proper infrastructure minimizing the communication delay between market and traders is crucial in allowing execution of data receipt, order submissions, and confirmations at the highest possible speed. Further, reading the full order book in real time and designing the more appropriate trading strategy is one of the key drivers of a successful trading strategy. Latency minimization goals are generally obtained by co-locating servers of market participants directly close to the market's infrastructure.

Regulation results in an increased role of algorithmic trader. In many cases, laws encourage market fragmentation and, at the same time, competition among market venues. In this case, the availability of tools that allow traders to place or cancel orders on multiple venues is essential to the ability to generate profits from arbitrage as a result of the lowest possible trading fees. Gomber et al. (2011) point out that regulatory changes are one of the key drivers of the recent innovations in financial markets.

This paper provides an overview of the most recent advances in trading systems and strategies, together with the cost of benefits of their respective implementations. The topics discussed in this chapter are presented in a sequential order. The next section presents the various definitions of AT, HFT, qualifying their importance in the current evolution of trading technologies. Next, the reader is introduced in the various order types: to appreciate trading execution techniques, the third section presents the definitions of basic limit and market orders together the most common instructions currently adopted in the trading practice (Fill or Kill, Immediate or Cancel, and so on). Having established the general terminology, a special section is dedicated to the description of the main algorithms, classified according to impact driven, cost-driven and opportunistic: within this context we will analyze the main characteristics of Volume Weighted Average Price (VWAP), Percent of Volume (POV), Time Weighted Average Price (TWAP) algorithms. The choice of execution is conditional to pre and post-trade analysis needed to estimate transaction costs involved in trading activity: this is discussed in a separate section where all the types of transaction costs both explicit and implicit are presented, following and extending the concept of Implementation Shortfall by Perold (1988). The choice of a trading strategy can be presented as an optimization problem: this is discussed in the section entitled "Optimal Execution Strategies", where it is presented the approach leading to the trading path which minimizes transaction costs, together with the concept of Efficient Trading Frontier highlighted by Almgren and Chriss (2000). A final step is about the practical implementation of the concept discussed. In a final section the Execution Tactics divided in Impact-driven, Price Risk driven and Opportunistic are presented with a special emphasis to hidden and iceberg order. The final section concludes. Basically, the first two sections introduce the reader to the main language and definitions. Then the next step is a general descriptions of the algorithms. To be implement efficiently an algorithm we need to understand the constraints: these are basically given by transaction costs. The design of an optimal trading strategy delivers a set of alternatives classified according to trade aggressiveness: this originates the Efficient Trading Frontier. The path for our presentation starts from definitions to bring the reader to the construction of optimal and implementable trading strategies.

## 2. TRADING ALGORITHMS CLASSIFICATION

The first problem when exploring the subjects of AT is the lack of a consistent set of definitions regarding algorithmic processes. However, Gomber et al. (2011) contends that HFT may be viewed as a subgroup of AT as professional traders observe market parameters or other information in real time and properly design specific trading strategies aimed at achieving specific goals without human intervention. These goals generally apply to DMA or SA technologies for order routing.

The main characteristics of AT include the following: (1) automated order submission; (2) automated order management; and (3) use of DMA. In contrast, the main characteristics of HFT include the following: (1) the ability to handle a large number of orders; (2) rapid order cancellation; (3) proprietary trading; (4 no significant position at the end of day; (5) a very short holding period and low latency requirement; and (6) co-location/proximity services. According to the Securities and Exchange Commission (SEC), HFT refers to "professional traders acting in a proprietary capacity that engages in strategies that generate a large number of trades on a daily basis." Brogaard (2010) estimates that HFT accounts for more than 50 percent of the overall daily volume on equity markets.

With the massive use of computerized trading system, market making has experienced further evolution: traders simultaneously quote buy and sell limit orders to profit from bid-ask differences. In some circumstances, this evolution is specifically imposed by market regulation, such as for the Designated Market Maker at the New York Stock Exchange (NYSE) or the Designated Sponsor via the trading system XETRA at the Frankfurt Stock Exchange. Basically, this feature highlights market making with obligations for registered market makers. Therefore, a close link exists between market making and HFT: designated liquidity provisions are a specific subset of rules of the market making strategies, which can be very efficiently implemented by using HFT. At the same time, non-mandatory market making strategies and not necessarily linked to liquidity provisions can be implemented with HFT. Market fragmentation allows traders to choose the best trading venue in terms of price discovery mechanism and liquidity.

Quantitative Portfolio Management (QPM) refers to the technical ability of portfolio managers to relate stock movements to other market data. This approach has the main goal of monitoring large securities portfolios for a prolonged period of time, as opposed to HFT where positions tend to be liquidated overnight. Greater human intervention can occur by using QPM. The strategy uses algorithms to generate trading decisions based on statistical calculations and data analysis techniques. QPMs are built in order to make automated analysis and decisions related to portfolio selection. The generation of trading signals is a consequence of previous analysis. This process can be automated as well, but the human role intervenes in the verification of the analysis and the trading strategies to be implemented. In principle, QPM differs from HFT since the latter does not have the goal of portfolio optimization and selection. On the other hand, the increasing role of latency and computer speed makes the distinction between HFT and QPM more evanescent.

An interesting aspect emerges as a consequence of both increased fragmentation and transaction in the role of Smart Order Routing (SOR) systems. SOR systems allow access simultaneously to several venues or liquidity pools to search for the best order destination to optimize order execution to minimize implicit and explicit trading costs. In practice, SORs scan predefined markets in terms of bid/ask quotes in real time and direct the order in a dynamic way toward the best venue offering more convenient conditions at the time of order entry. This process is an extremely complicated because achieving the best results requires disposing of a routing system able to screen in real time the entire order flow/order book composition of each trading venue in the market together with the price evolution and to compute the total execution costs of each trades in each different market, by applying the explicit cost dimension (commissions, fees and taxes). This activity is mainly reserved for fully automated systems for HFT, which have the ability to reduce search costs and increase search speed across trading venues.

Overall the assessment of AT and HFT misses the fact that both procedures represent one of the most important advances in modern market arena. Given the large complexity of actions traders must implement (risk exposure analysis, gross position netting, matching order size and strategies with regulations and market rules and so on), algorithms can be of great help in mitigating cognitive limits associated with the human mind and behavior in collecting and processing large masses of information in a few seconds. Another benefit of algorithms is their ability to provide liquidity when it is scarce.

Biais, Foucault, and Moinas (2010) maintain that there is a problem of adverse selection after the introduction of algorithms. Asymmetric information emerges between slow and fast traders: fast traders observe information before other agents. Because fast traders can profit from such knowledge, slow traders lose. The authors show that the increase in algorithm trading might cause an increase of adverse selection costs when slow traders are evicted from the market. Empirically, Jovanovic and Menkveld (2010) have verified this when fast algorithm trader enters Chi-X (a Dark Pool listed in the Dutch Market). A fast drop in volume is clearly documented, probably due to slow traders pulling away from the market. The risk with AT and HFT is then given by a multiplicity of equilibria where identifying positive and negative effects is possible. At the moment, no strong evidence either favors or opposes algorithms.

## 3. ORDER TYPES AND TRADING STRATEGIES

A distinction exits between two types of orders in the market: limit orders and market orders. With a *limit order*, a trader specifies and orders to buy (or sell) a quantity of securities at a pre-specified price. A *market order* is an instruction to trade a given quantity at the best possible price. The main difference between the two types of orders involves the immediacy need: the market order is a demand of immediacy expressed by a trader because it does not impose a price limit on the transaction. As a result, market orders have a high probability of being executed. Uncertainty surrounding this type of order relates to the execution price. Conversely, a limit order fixes a price limit above or below which the order should not be executed. For a buy limit order, the execution must occur at or below the limit price; while for sell limit order, the execution must take place at or above the price limit. Limit orders are more uncertain than market orders because if the book does not contain a proper match, the order either will expire or remain unexecuted. If the order is partially executed, a residual quantity can be left on the order book.

Additional instructions can be attached to the main order. Some of the most common features include the following:

- *Fill or Kill (FOK)*: A fill or kill instruction ensures that either the order gets executed immediately in full or not at all.
- *Immediate or Cancel (IOC)*: An immediate or cancel instruction states that any portion of the order that cannot be executed immediately against any order on the book is cancelled. IOC orders can be associated with both market orders and limit orders. When a limit price is inserted, the order will execute as the required quantity is available.
- *All or none (AON)*: This instruction forces a 100 percent completion requirement on an order. The focus after inserting this instruction in not on immediacy. In fact, there might be some persistence of the order on the book.

According to the time duration of presence on the order book, orders are classified in the following manner:

- *GTD, Good Til Date*: The order remains active until the end of the trading session at the pre-specified date. For example: GTW, for Good This Week; GTM, for Good This Month.
- *GTC, Good Til Cancel*: The order should stay active until the user specifically cancels it (Good Till Expiration).
- *GAT, Good After Time Date*: The order allows control over when it becomes active. This type of order is useful for broker systems

where clients may choose to spread orders over the day. In trading algorithms, this instruction is a start-time parameter initializing the trading process.

Among the different order types, various preferencing and directed instructions may exist. Such instructions allow directing orders toward a specific broker or dealer or trading venue. The order can contain the routing instructions including the price and quantity instructions.

Some specific routing instructions are:

- *Order protection rules in the US*: Brokers and venues are responsible for finding the best price. The host venue should either reject or forward the order to the market when achieving a better price is possible.
- *Additional instructions*: Such instructions as Do not Route or Inter-Market Sweep allow for better control on the order routing.

A specific and interesting order instruction is represented by flash orders. These orders display the source venue for a short period before being routed to another destination. The idea underlying such orders involves the possibility offered to a trader of achieving the best price. Only AT and HFT can exploit the advantage emerging from flash orders because technological requirements are binding. These types of orders are under scrutiny because of the advantage assigned to specific traders. In some senses, the asymmetric information phenomena previously discussed becomes part of the problem.

Hybrid order mechanisms combine features typical of limit orders together with those of market orders. An example of this combination is given by market with protection orders, whose goal is to offer the immediate execution of a market order with the protection of a price limit. In this type of order, the limit price is established as being different from the last execution price. This type of order balances the certainty of execution and certainty of price.

Sometimes, specifying the conditions for the execution of a given order may be useful. Under these conditions, conditional order types emerge. Such orders are valid under a given set of conditions, usually determined by the limit assigned to the market price. If the specific conditions (say, with respect to the price limit) are met, then the order is effectively executed.

*Stop orders* (sometimes also called *stop loss orders*) become binding when the market price reaches a given threshold. In this case, they are transformed into market orders. In continuous trading, the triggering price is the last traded price. In an open outcry auction, the triggering price is the closing price. Activation occurs for buy orders when the market price hits the stop price or above. For sell orders, activation occurs when the market price goes below the threshold. For example, suppose an investor initiated a long position at a price of 10. It the price increases to 11, a stop sell order is placed at 11 as a means of protection. If the price trend reverts, the stop threshold will be triggered. Alternatively, the stop order is never implemented. Such order types are implemented and combined with the market with protection order types in order to increase the probability of execution and to obtain strong price controls.

Next, this paper discusses hidden orders and iceberg orders. These types of orders are at the core of the functioning of some specific venues, such as the "dark pools," where the identity of traders is hidden alternatively or jointly with the full size of the order book. Transparency in financial markets allows traders to correctly track demand and supply. Conversely, when some traders need to exchange large size orders, consequences may be too strong based on the price patterns. For example, a large order might convey wrong information to the whole market and might generate herd behavior, which, in turn, could cause strong oscillations. For these reasons, many trading venues often allow the retention of large orders hidden in the book. Moreover, a compromise between hidden orders and visible orders is represented by iceberg orders, where only a fraction of the order is made visible to the market. If the market allows hidden orders, traders can participate without revealing their positions to the market. To further illustrate, consider the example reported in Table 1 that presents the order book for trades of a given stock.

(Insert Table 1 about here)

Currently, a market order is hidden to purchase 2,000 shares of stock. The book outlined in Table 1 represents the situation before the order is sent to the market. If the order is kept hidden, after the order has been routed to the venue and executed, the book after the execution can be represented to the right, where the depiction of the effects on the sell side of the book with 1000 shares being matched with the initial order. Note that the order does not appear on the buy side since it is a hidden order. As a market order, it does not have price priority and can immediately be matched with the best prices offered by the sell side. Secondly, after having executed half of the order (1,000 shares), the remaining part becomes latent, standing on the book. If an IOC is issued to sell additional 1,000 shares at 51, the investor can "ping" the market to attempt to cross with the remaining portion of the order. If an order to sell occurs, and the order is executed, the implication is that on the book there was a hidden order ready to be exchanged.

A different pattern can be observed for iceberg orders. In this case, only a small part is visible, with the remaining portion of the order hidden. The visible portion of the order is virtually non evident from other types of limit orders. When the visible portion of the order is executed, the trading system slices a new order from the hidden part until the whole order is completed. The visible part has time priority, while the hidden portion has only price priority.

(Insert Table 2 about here)

Consider the example of the iceberg order from Table 2. In the table on the left side, a portion of the sell order S1 has been issued on the book, the whole order is H1, which is not visible. When order S1 at a price of 51 is executed, a new portion of the hidden order is released from the hidden order H1 and is included in the book as order S4. If there were two hidden orders at the same price, time priority indicates which order gets executed first.

The current market practice has created a more granular structure of orders. In the present context, this paper limits its treatment to the most important. The next section discusses the main trading strategies and related algorithm selection.

## 4. ALGORITHM SELECTION

As discussed previously, an algorithm identifies a set of instructions to accomplish a specific trading task. Three types of algorithm are described below:

- *Impact driven algorithms*: These algorithms attempt to minimize the overall market impact. They are established in order to reduce the effect that trading activity might have on the market price.
- *Cost-driven algorithms*: The underlying goal of a cost-driven algorithm is cost minimization. These algorithms do not take into account market impact, timing risk, and other related factors such as the measures of benchmarking and implementation shortfall.
- *Opportunistic algorithms*: These algorithms take advantage of market evolutions

### 4.1 Impact-driven Algorithms

The most common impact algorithms are named time weighted average price (TWAP) and volume weighted average price (VWAP). The general feature of such algorithms is that they focus of the behavior of a pre-specified benchmark and track it systematically with limited sensitivity to market conditions.

*TWAP*

A example of TWAP is represented by the strategy to buy 10,000 shares of a given asset. Two strategies are possible. First, a trader could issue regular-sized orders to buy 500 shares every 15 minutes for five hours. A second strategy can be implemented such that the trader issues an order to buy 1,000 shares every 15 minutes for 2.5 hours. The trade profile is very simple and can be easily forecasted, implying a signalling risk.

After systematically observing the behavior exhibited with such trades, the market can easily forecast the next trade and implement actions that may create adverse price movements. After the trading pattern is discovered, the only unknown to the market is the total size of the order. An alternative to this behavior might consist of inserting a tilting factor, acceleration or deceleration of the order size, after an established threshold of the total order has been fulfilled.

As an example, the trader may insert a tilting factor that results in the order executing more quickly after 30 percent of the total order has been executed. Another alternative is the insertion of a randomized approach that can be adjusted according to the position the fulfilment of the order has been completed. In general, the tilting factor defines the degree of aggressiveness attached to the algorithm. An aggressive strategy occurs when a larger percentage of the order is executed earlier. In contrast, a passive strategy occurs when the order size is smoothed out over time. The algorithm is generally completed by the presence of some parameters, such as an on/off switch that highlights how far the order might be from the benchmark. Another choice parameter is the frequency of trades and the degree of randomization of the trade sizes.

The more aggressive the order strategy, the greater is the price impact associated with the trade. The primary shortcoming of the TWAP strategy is associated with a passive strategy. An alternative to the passive strategy involves implementing a trading schedule with a dynamical adjustment path that evolves according to the general market conditions.

*VWAP*

The main property of VWAP is its ability to design trading strategies according to the total market volume. More generally, the definition of the VWAP is given by Equation (1):

$$VWAP = \frac{\sum_{i=1}^{n} V_i P_i}{\sum_{i=1}^{n} V_i} \tag{1}$$

where $V_i$ is the size of the $i$th trade and $P_i$ is the price associated to the $i$th trade. VWAP identifies the ratio of total traded value to the total traded quantity. In Equation (1), $n$ represents the number of trades. The main way in which VWAP differs from TWAP is by the $n$ variable: the total number of trades when the strategy is formulated can be unknown. Estimation methods on the basis of historical volume can be used to solve for the optimal number of trades. Kissel and Glantz (2003) contend that the optimal trading strategy can be found by using Equation (2):

$$X_j = z_j X \tag{2}$$

where $z_j$ is the percentage of daily volume traded; X is the total volume traded; and $X_j$ is the target quantity. Thus, VWAP can be re-interpreted as shown in Equation (3):

$$VWAP = \sum_{j=1}^{n} z_j \bar{P}_j \tag{19.3}$$

where $P_j$ is the average price level in each period. According to Equation (3), VWAP is defined as the weighted average value of the prices associated with trades executed to implement a given strategy.

The VWAP trading activity is dependent on the historical pattern of the trading volume, not from actual volume or actual price changes. The performance evaluation of VWAP is given by how closely it tracks the target and predicted market values. If historical and actual market values do not differ substantially, then the performance of the VWAP strategy will not suffer. The performance of the VWAP strategy is less effective if there are substantial differences between historical and actual traded volume.

VWAP can be complemented by special tracking parameters that monitor how closely the algorithm tracks the target together with trending or tilting instructions. The objective of the VWAP strategy is to adjust the trade path to closely target the execution method.

*Percent of Volume (POV)*

POV is an algorithm designed to closely target the market volume. In fact, with TWAP and VWAP, the trading schedule is deterministically determined, independent on the market volume. Conversely, the trading schedule based on POV is dynamically adjusted. The deterministic choice involves the proportion of market volume, or participation rate, that is to be targeted  For example, by establishing a 10 percent participation rate, the POV algorithm implies an order submission strategy equal to 10 percent of the total market volume (or possibly associated with either the buy or sell side of the order book). As a result, the trader no longer has any certainty about the trade completion within a specific time period because trade size is dependent on total market volume. Although POV is similar to VWAP, the difference is given by the actual volume employed by POV as opposite to historical or estimated volume employed by VWAP. In practice, POV reacts to repeated trades i trading target with the observed value. Clearly, the trade sizes need to be adjusted to take into account the own trading. The adjustment factor is given by *1/ (1-pr)* where *pr* is the participation rate.

One of the main problems associated with this algorithm is given by the competition for liquidity. If several traders adopt a similar algorithm for trading illiquid assets, the direct consequence can be a strong price pressure as liquidity is reduced by absorbing available market volume. Thus, using this algorithm is often accompanied by price limits associated with excessive liquidity tension.

To avoid the risk of the market being able to predict the trade schedule implemented with POV, order aggressiveness and placement is often adjusted with tilting deterministic parameters. Further, some safeguards are often inserted in order to prevent excessive fluctuation of market volume. This is usually obtained by inserting the maximum trade size.

A variant of POV is offered by its price-adaptive version. In this case, the participation rate is adjusted according to how the market price compares to a benchmark, which could be either a market index or an exchanged traded fund- (ETF) based index. Trade aggressiveness is reduced when price tension is high and relaxed if price tension is low. Another modification of the POV strategy can be implemented when another type of price condition is added to the algorithm such that child orders are not going to receive a price higher than the last trade or the current best bid.

Overall, the dynamical features of POV make this algorithm more attractive and less forecastable than the simple VWAP or TWAP. However, regulatory authorities are now leading market participants to adopt variants of POV that may not endanger the safety of the market as a whole. In this case, including as such variants as an adjustment mechanism towards a market benchmark with mandatory parameters may be possible. Currently, no formal prescriptions made by SEC, FSA or other regulatory agencies exist so more research needed in this field.

**4.2 Opportunistic Algorithms**
Opportunistic algorithms are designed to take advantage of favorable market conditions. Liquidity-driven algorithms designed to search for liquidity across several market venues play a key role in this process. After the introduction of algorithms, liquidity has become even more central in judging the quality of market status. Traditionally, liquidity has been associated with lower bid-ask spreads, focusing exclusively on the explicit transaction costs. As this paper will clarify after the presentation of transaction costs analysis, a non-liquid asset implies excessive price volatility after a trade, signalling a trading activity even slightly above what the market may expect. The increasing degree of fragmentation occurring in modern markets delivers a more careful definition of opportunistic algorithms. In this context, liquidity-driven opportunistic algorithms are designed to search for liquidity over multiple execution venues. The most important element is given by liquidity aggregation across venues with technology that is able to collect order book data from all possible venues for the purpose of collecting everything into a unique order book. The design of a proper opportunistic algorithm should take into account the following: (1) the fee structure associated to each trading venue; (2) the latency, which in this case represents the time lag between orders sent and processed from each single venue; and (3) the probability of execution associated to each trading venue. Orders are aggregated by using price priority and then by the probability of execution.

Given that algorithms are designed to capture the performance of the entire order book, implementing a way to track the behavior of the order book over time is extremely important. In this context, historical data analysis and a close monitoring scheme of the order book may help in estimating hidden liquidity. Often, the setup of the opportunistic algorithm includes specific actions such as "pinging for liquidity" in the book of a given trading venue. This practice consists of inserting a specific type of order like IOC or FOK to test for liquidity inside the order book. If an IOC or FOK gets executed, then hidden liquidity exists on the book.

In practice, opportunistic algorithms represent a form of liquidity-seeking strategies that do not react to traded volume. Instead, it is the market depth, reflecting the volume at each favorable price point that represents the indicator leading the strategies of the algorithm. When prices are favorable, the algorithm is designed to trade more aggressively to consume liquidity. When market conditions are unfavorable, participation trade falls to zero. Of course, in this context, liquidity plays a crucial role. With such types of algorithms, orders do not necessarily complete. The algorithm often contains a set of warnings about the liquidity situation of a given security in each trading venue. Some of the key parameters to be specified at the order initiation include visibility and benchmark practice. Visibility specifies how much of the order may actually be displayed at each execution venue. For example, in the case of low visibility, IOC order types are largely employed. An alternative to this is represented by processing the order through dark pools. The role of benchmark pricing is employed to decide if the security's price is favorable enough to measure participation.

## 5. PRE-TRADE AND POST-TRADE ANALYSIS

In trading activity, the role of transactions is crucial and an accurate quantification of them helps traders to properly decide when and where to make transactions. In this context, pre-trade analysis is very important because it helps traders to make informal decisions about the best order execution. Pre-trade analysis is conducted by using information coming from data. There are two sources of information employed for this type of analysis. The first source comes from information relative to fundamentals of securities such as price/earnings for stocks and coupons for fixed income. The second source emerges from trade-related sources: prices, liquidity, and cost estimates.

Information is gathered from trade-related sources relative to price ranges, trends, momentum. Obtaining liquidity information permits getting average daily volume (ADV), volume profile, and indicators of trading stability. Risk indicators include volatility, beta, and risk exposure. Cost estimates are defined by measures of market impact and timing risk. The estimate of transaction costs implies exactly the importance and the difficulty of executing a given order.

Post-trade analysis delivers results about broker/trader performance for both investment and execution decisions. The key element of the post-trade analysis is obtained by breaking down costs into several components including fees and commissions. With post-trade analysis, establishing the relative performance, net of costs, is possible for the trading activity implemented by a trader. Within this context, the choice of the benchmark becomes essential in performance computations. The average execution price of any given trades has to be compared with a benchmark. Post-trade benchmarks are based on closing prices. The main purpose of these benchmarks is to determine the mark-to-market computation and profit and loss. The problem with closing prices is their scarce representative power of real market conditions. An algorithm can be designed in order to track closing prices. However, closing prices tend to be more sensitive to the order flow than in other periods during the day. This situation implies that orders can be more exposed to timing risk and to more volatile conditions.

Intra-day benchmarks are generally given in the following manner: (1) the average of open-high, low-close, (2) TWAP and (3) VWAP. The average of the open-high represents an average over only four data points and can be easily distorted by extreme values TWAP represents an average of all observed trade price over a given timing period. Thus, all trades are weighted equally, which has the drawback that smaller trades occurring at extreme prices may markedly affect TWAP calculations. However, TWAP is generally employed when trade volume data are unavailable. In general, VWAP is probably the fairest indication of how market prices have changed over a specific time span. VWAP does not have a strong impact on small trades and can be considered as a very good representation of average market conditions. A higher value of VWAP indicates a higher market impact determined by the last trade. Conversely, VWAP does not deliver good performance when the sample of analysis includes large order or when the volatility of orders is large. At the same time, market conditions that include large price reversals do not represent an ideal context to apply VWAP.

**5.1 Pre-trade analysis**
Pre-trade benchmarks can be distinguished into those that are directly available and those that not publicly observable. Directly available benchmarks are represented by the previous close and opening price. Such benchmarks do not necessarily reflect actual market conditions during the trading day. Other useful prices include the decision price and arrival price. The *decision price* is the price at which the choice to invest has been actually made. The decision price is used in the calculations related to implementation shortfall, which will be discussed in a later section. The *arrival price* is the price registered at the time when the order was actually traded, which is the time at which the order arrives from the investor. The problem with such information is that the decision/arrival price is not always recorded by the investor.

Kissel and Glantz (2003) propose a Relative Performance Measure (RPM) based on the comparison of what the trade achieved in relation to market condition. Equation (4) presents RPM for volume while Equation (5) gives the RPM for trades.

$$RPM(volume) = \frac{\text{Total Volume at prices less favorable than execution}}{\text{Total Market Volume}} \quad (4)$$

$$RPM(trades) = \frac{\text{Number of Trades at prices less favorable than execution}}{\text{Total Number of Trades}} \quad (5)$$

The advantage of RPM is that it is already normalized in percentage terms. For example, a trade achieving 90 percent of RPM is better than another trade reaching 60 percent.

**5.2 Post Trade Transaction Costs**

The most important measure of performance in post-trade analysis is *implementation shortfall* (IS), which is the difference between the price that an investor decides to trade and the average execution price that is actually achieved. The decision price is the benchmark. If the decision price is not specified, the mid price is generally accepted as default benchmark. Perold's (1988) measure of IS consists of comparing the return on a "paper" (hypothetical) portfolio with the return obtained in the real, actually traded portfolio as shown in Equation (6).

$$IS = Return_{paper} - Return_{real} \quad (6)$$

The performance of the theoretical or paper portfolio depends on three factors: (1) the price where the decision to invest was made, $P_d$; (2) the final market price $P_N$; and (3) the size of the intended investment $X$. More formally, given $x_j$ the size of the j$^{th}$ order, with achieved price $p_j$, the IS can be defined as shown in Equation (7):

$$IS = X(P_N - P_d) - \left(XP_N - \sum_j x_j p_j - \text{fixed}\right) \quad (7)$$

Assuming that the order is fully executed, the IS measure, as shown in Equation (8), becomes:

$$IS = \sum_j x_j p_j - XP_d + \text{fixed} \quad (8)$$

If the order is not fully executed, the IS can be rewritten as shown in Equation (9):

$$IS = \underbrace{\sum_j x_j p_j - \left(\sum_j x_j\right)P_d}_{\text{Execution Cost}} + \underbrace{\left(X - \sum_j x_j\right)(P_N - P_d)}_{\text{Opportunity Cost}} + \text{fixed} \quad (9)$$

where $X - \sum_j x_j$ indicates the size of the unexecuted position.

According to Kissel and Glantz (2003), transaction costs can be classified in two ways: (1) fixed and variable and (2) visible and transparent. Typically, visible costs cannot be managed, while non-transparent components are manageable through an appropriate execution strategy.

The Perold's IS measure can be expanded by adding investment-related delay costs, as suggested by Wagner and Glass (2001). In this case, the expanded measure of transaction costs is given in Equation (10):

$$TC = \underbrace{\sum_{j=1}^{n} x_j (P_d - P_0)}_{\text{delay cost}} + \underbrace{\sum_{j=1}^{n} x_j P_j - \sum_{j=1}^{n} x_j P_0}_{\text{Trade Related}} + \underbrace{(X - \sum_j x_j)(P_n - P_d)}_{\text{Opportunity Costs}} \qquad (10)$$

where X indicates the total amount of shares traded at $j^{th}$ transaction; $P_j$ is the price of the $j^{th}$ transaction; $P_d$ is the price observed at the time of investment decision; $P_0$ is the price at time when the order is released to the market; $P_n$ is the price at the end of trading session; and *n* is the total number of trading periods.

Cost minimization strategies focus on developing analytical methods to manage trade-related transaction costs. Minimization of delay costs and opportunity costs is a natural consequence of adopting a fast trading method, with the ability of minimizing trade latency as much as possible. The rapid increase of HFT techniques and infrastructures is motivated by the need to minimize delay and opportunity costs. The design of an optimal execution method allows reducing the trade-related component of transaction costs.

## 6. OPTIMAL EXECUTION STRATEGIES

Trading activity is a complex set of strategies, market conditions, and costs. Trading too aggressively produces a higher market impact. Conversely, trading too passively implies a higher timing risk. In general, optimal execution strategies attempt to find a balance between these two extremes. Synthetically, the optimal execution strategies can be represented by Figure 1.

(Insert Figure 1 about here)

The market impact is decreasing over time by trading strategies, while timing risk increases over time. The optimal execution strategies mix up the balance between these two cost measures. In general, the optimal solution is determined by the trader's degree of risk aversion. The balance between market impact and timing risk defines the famous optimal trading horizon.

The design of an optimal algorithm has the goal of minimizing total costs related to trading activity. The characteristics of a transparent and efficient algorithm are given by two elements: (1) the choice of a desired benchmark price (close, open, and arrival) and (2) the choice of a desired trading style: aggressive, normal, moderate, including the choice of market adaptation (representing the choice about how the algorithm should adapt to market conditions). Algorithms should built in order to balance the tradeoff between market impact and timing risk over the trading horizon. According to Kissel and Glantz (2003), market impact is the movement in stock caused by the order or the trade. Permanent impact costs are the result of an information leakage. Almgren and Chriss (2000) focus on designing optimal execution strategies. The difficulty with their approach is the need to determine an approximate market impact function at the trade level that also takes into account price movements. Kissel and Malamut (2006) solve this problem by considering a top-down approach, where the average market impact cost is estimated for aggregated trade imbalance and allocated to trading periods.

According to Kissel and Malamut (2006), the impact cost is defined according to Equation (11):

$$I = a_1 \left(\frac{X}{ADV}\right)^{a_2} \sigma^{a_3} X P_0 \tag{11}$$

where X is the number of shares contained in the order; ADV is the average daily trading volume; σ is the annualized volatility; and $P_0$ is the current price. The market impact function is given in Equation (12):

$$MI(x_k) = \underbrace{\sum_{j=1}^{n} \frac{b_1 I x_j^2}{X(x_j + 0.5 v_j)}}_{\text{Temporary Impact}} + \underbrace{(1 - b_1) I}_{\text{Permanent Impact}} \tag{12}$$

where $x_j$ indicates the number of shares transacted in period ${}_j$; $v_j$ is the expected volume (net of the order) in period ${}_j$; and $b_1$ is the percentage of temporary market impact costs. Assuming the existence of a constant trading rate, Equation (13) emerges in the following manner:

$$\alpha = \frac{x_j}{v_j} \tag{13}$$

After rearrangement (Kissel and Malamut, 2006), the entire expression can be rewritten for the market impact given in Equation (12) as a function of the trading rate shown in Equation (14):

$$MI(\alpha) = \underbrace{\left(\frac{b_1 I}{X}\right)\left(\frac{2}{3}\right) \alpha^{1/2}}_{I_1} + \underbrace{\frac{(1-b_1)I}{X}}_{I_2} \tag{14}$$

which, after rearrangement, becomes Equation (15):

$$MI(\alpha) = I_1 \alpha^{1/2} + I_2 \tag{15}$$

Equation (16) provides an alternative formulation for market impact:

$$MI = \underbrace{I(Q, \sigma)}_{I_{bp}} \underbrace{[\mu^{-1} b_1 + (1 - b_1)]}_{d(\mu)} \tag{16}$$

where $I(Q,\sigma)$ is the trade impact function whose arguments are market imbalance Q and volatility σ. Function $d(\mu)$ is the dissipation function, where $\mu = V_{\text{side}}/Q$ and $V_{\text{side}} = \sum_{i \in Q} \text{sign}(v_j)$ indicates the cumulative amount of order on the same side of the market (in buy or in sell, according to the sign of $v_j$).

Given the above information, the problem of optimal trade execution consists in finding the optimal trading rate that minimizes the cost, given the risk. Formally, the problem can be stated in Equation (17):

$$\begin{cases} Min\ \phi(x_k) \\ s.t.\ \ \mathcal{R}(x_k) \leq \mathcal{R}^* \end{cases} \quad (17)$$

where $\mathcal{R}^*$ is the threshold level of timing risk. Following Kissel and Glantz (2003), the timing risk is the uncertainty surrounding the expected market impact cost that is related with volume and intraday volume patterns. By definition, timing risk is given in Equation (18):

$$\mathcal{R}(x_k) = P_0 \sqrt{\sum_{j=1}^{n} r_j^2 \frac{t\sigma^2}{n}} \quad (18)$$

In terms of the trading rate *a*, the timing risk can be reformulated as shown in Equation (19):

$$\mathcal{R}(\alpha) = P_0 X \sigma \sqrt{\frac{s}{3\alpha}} \quad (19)$$

The goal of finding optimal execution consists of finding the trading rate *a* which minimizes the expression in Equation (20):

$$\min_\alpha \{MI(\alpha) + \lambda \mathcal{R}(\alpha)\} \quad (20)$$

A crucial parameter in Equation (20) is $\lambda$, which represents the degree of risk aversion of the trader. In other words, choosing the value of $\lambda$ can affect the aggressiveness of the trading strategy. A lower $\lambda$ indicates a patient trader because less weight is assigned to timing risk and more weight is assigned to trading cost. Conversely, a higher $\lambda$ implies a more aggressive trading strategy with higher weights assigned to market impact costs and lower weights assigned to timing risk.

Figure 1 captures a pictorial representation of the optimization procedure. The figure illustrates the evolution of market impact cost and timing risk as a function of trading rates. Clearly, the optimal trading rate is obtained as the minimum point of the curve outlined by Equation (20), for any given level of risk aversion $\lambda$. Given the optimal strategy, which is defined in terms of the trading rate as a function of the parameters and exogenous variables of the model, Equation (20), conditional to the optimal solution, delivers the efficient trading frontier (ETF) as defined by Almgren and Chriss (2000). In Figure 2, the ETF is shown for two given levels of price benchmarks adopted to study trade performance.

(Insert Figure 2 about here)

The dotted-line ETF has been constructed adopting the previous close price as a benchmark, while the solid-line ETF is based on the arrival price. In general the ETF computed with previous close price can be above or below ETF conditional to arrival price for two reasons. First, the ETF includes permanent price impacts, not just the temporary component. Second, the ETF includes price movements between the time of investment decisions and the trade starting time.

By definition, the ETF is the locus of all optimal trading strategies, i.e., the least cost for a given level of timing risk and the lowest timing risk for a specified cost. An efficient strategy is on the ETF. All strategies above the ETF are inefficient, in the sense that finding another strategy with lower timing risk, given the expected cost and/or a lower expected cost, given the timing risk, is always possible. Movements along the curve are governed by the size of parameter $\lambda$. That is, if $\lambda$ is higher (implying a more aggressive trading strategy), the optimal trading point will be located in the top left part of the curve, while for lower $\lambda$, the solution will be associated with lower expected cost and higher timing risk.

The notion of the ETF has recently expanded to several other optimal trading models that include the uncertainty of prices. Almgren (2012) develops an optimal trading model with an arithmetical Brownian motion process for the stock prices. Gatheral and Schied (2011) expand their optimal trading model on the basis of geometric Brownian motion. This area of research is rapidly expanding which is not surprising given the strong impact that this type of research will have in the design of trading strategies and on its global market impact.

## 7. EXECUTION TACTICS

Passive approaches adopt limit orders priced at or behind the market whereas aggressive approaches adopt either market or marketable orders. Neutral tactics tend to be more flexible. They start passively to seek price improvement. If they fail, these neutral tactics should be replaced with more aggressive ones. A more dynamic tactic strategy considers market conditions and adapts strategy to it. Broadly speaking, execution tactics may be classified on the basis of the goals that drive their usage:
- *Impact-driven tactics* seek to further reduce market impact by splitting the order into smaller quantities or by hiding a portion of it.
- *Price/risk driven* approaches strive to dynamically adjust based on market conditions.
- *Opportunistic tactics* seek out advantages when conditions are favorable.

### 7.1 Impact-driven Tactics
*Slicing*
Order slicing is the precursor to the early schedule-driven algorithms. Reducing the size of an order can lower its market impact and any associated signaling risk. Randomization is a way of further reducing signaling risk. This reduction applies to both quantities to be split and the time between each child order. When adopted sequentially, order slicing effectively acts as a hiding mechanism. The strategy may be adopted to create synthetic iceberg orders for venues that do not natively support this order type. For native iceberg orders, the hidden portion is still part of the order book so it may still participate in the trade crossing mechanism, with only price priority. Order slicing, instead, relies on execution confirmations to know when next to send another order slice. These orders are dispatched after a matching process is employed. In this way, a trader may miss some crossing opportunities. For example, consider a requirement to sell 10,000 shares of ABC stock from which the trader has split an initial child order for 1,000 shares at 51 as S1. The main advantage of order slicing is that the trader has much more control of the display size. In this case, the second order can be split with a random size, rather than simply splitting 1,000 shares each time. Order slicing can be also applied in parallel, allowing simultaneous trading across several execution venues. The example reported in Table 3 represents the effect of an order to buy 2,200 shares on the order book.

(Insert Table 3 about here)

In Panel B, note the highlighted effect due to MO crosses. In Panel C, note that for a synthetic iceberg order, an immediate cross exists for orders S1 and S2 at 51. The next cross will be with S3 at 52. After the venue sends confirmation for these fills, there will be another split for S6 for 850 shares. In Panel D, the table illustrates the effects from a native iceberg. In this case, the hidden size completes the market order at 51. A new display size would have been split as order S6, 400 shares of which complete the market order, the remaining 600 shares stay on the order book.

*Hiding*
Hiding is associated with reducing signaling risk, which is the potential information that trading pattern relays to other market participants. The example reported in Table 4 records large block orders.

(Insert Table 4 about here)

Algorithmic orders are implicitly hidden because most strategies only release a small portion of the order for immediate trading. Liquidity is another critical variable because even a small order may have important price effects in case of illiquid assets. Discretionary orders allow a trader to hide his actual limit price. In Panel A, the intention to execute an order S3 with a limit price 52 exists even though the trader was prepared to trade at 51. The trader could increase his discretionary amount to shift his order deeper in the order book. This change would place more orders in front of the trader in terms of execution priority. Iceberg orders allow the trader to display only a portion of the order, with the remaining portion retains price priority. For example in Panel B, 1,000 shares are displayed with the remaining 7,000 shares hidden until S3 is completed, although it will maintain price priority over S4. This example illustrates the importance of obtaining the right balance between the visible and hidden parts. Smaller trades may take too long to complete with larger visible portions having greater predictability.

**7.2 Price Risk Driven Tactics**
Price risk driven tactics are based on changes in the spread and short-term price trends. As the gap between the best bid and offer narrows, a trader can afford to pay the spread, and the tactics may indicate issuing more marketable limit orders or market orders. When the spread widens, more passive pricing is used. Market conditions are also important. In a favorable price trend, a passive trading style tends to price orders farther away from market to try to take maximum advantage of price improvements. Conversely, during unfavorable price trends, a passive strategy may price orders closer to the market in order to reduce potential losses. Aggressive price strategies tend to rely on trends mean reverting so that they behave in the opposite fashion.

Risk-driven tactics consider the asset's price volatility and how much time is left to execute. The tactics may be combined to create a timing factor, which can be incorporated into order placement decisions. For a liquid asset at the start of the execution, the timing factor is minor, but increases over time. For less liquid assets, the timing factor is more important. Other execution tactics are defined as catching, which is based on cutting the losses when the price looks to be trending away.

The layering tactic simultaneously maintains a range of standing limit orders. Orders are spread throughout the order book, usually with different limit prices. This approach attempts to take advantage of favorable price movements. With price/time priority, layering allows for the preservation of time priority for each order. This occurs because as the market price moves, a new order can be split, rather than just updating an existing order to match the new level. Time priority is very important for liquid stocks that have a densely populated order book. If time priority is lost, the only way to regain it is with a more aggressive price. In general, layering is useful for highly liquid assets with dense order books, where achieving time priority is difficult. In this way we can track price moves without significantly affecting our overall execution probability.

### 7.3 Opportunistic Tactics
These tactics tend to maximize the benefits from favorable market conditions, such as liquidity. Three types of such tactics follow.
- **Seeking**: This tactic aims to source additional liquidity from hidden orders. Seeking tactics may also focus on reducing signaling risk. Monitoring the state of the order book can enable creating models to estimate the probability of how much volume is hidden at the various price levels throughout the order book. Another important concern when searching for liquidity is to keep orders hidden. The goal is to fill the orders at the best price possible without giving away information about the actual requirements. The choice is between various fill instructions available for limit orders. Only IOC or FOK are suitable because the others all leave residual orders on the order book. IOC allows partial fills whereas FOK is a strict execution.
- **Sniping**: This tactic allows the capture of liquidity while minimizing signaling risk. The tactic is a way for liquidity demanding traders to hide their strategy. As such, sniping is a variation of the seeking tactic for visible liquidity. To reduce signaling risk, traders use marketable limit orders with specific fill instructions. When liquidity becomes available, an aggressive order is used to cross with it.
- **Routing**: This tactic chooses the best destinations to send orders. Many factors have to be analyzed before deciding where to send an order including: (1) the probability of hidden liquidity; (2) the likelihood of successful execution; () latency; and (4) the venue specific costs or fees. Routing mechanisms make these decisions, based on available market data. Liquidity aggregation is the key aspect for trading in a fragmented market. Virtual order books can be formed by collecting data from all possible execution venues.

## 8. SUMMARY AND CONCLUSIONS

This paper presents the key piece of knowledge to follow based on the most recent advances in market evolutions, mainly represented by HFT and AT. The order characteristics and all the variants of the basic limit and market order types are considered for different trading needs. Given that AT and HFT involve using a specific trading strategy, the paper also presents the main characteristics of the most often employed trading strategies such as VWAP, TWAP, POV and others. Implicit and explicit transaction costs are also introduced, which are at the core of the choice process of an optimal trading strategy. Also introduced is the optimal trade execution of Kissell and Malamut (2006) together with a full description of the EFT.

The material covered in this paper is evolving quickly and is new to the microstructure literature. The authors believe that adopting HFT and AT will radically modify the way in which researchers and practitioners approach the field of market microstructure and market architecture. Many research areas are now open to these new advances, which challenge traditional concepts about market quality, liquidity, and price discovery. HFT has a role in propagating liquidity shortages and generating market crashes. While limited research exists on this topic, the design of optimal execution methods that do not endanger market conditions will be the challenge for the future.

**Table 1 Hidden Order Example**

This table contains the behavior of the order book when hidden orders are implemented.

| BUYS | | | | SELLS – before | | | | SELLS - after | | | |
|---|---|---|---|---|---|---|---|---|---|---|---|
| Id | Time | Size | Price | Price | Size | Time | Id | Price | Size | Time | Id |
| B1 | 10:41:00 | 2000 | 50 | 51 | 1000 | 10:35:00 | S1 | ~~51~~ | ~~1000~~ | ~~10:35:00~~ | ~~S1~~ |
| B2 | 10:35:10 | 1000 | 49 | 52 | 1500 | 10:37:15 | S2 | 52 | 1500 | 10:37:15 | S2 |
| B3 | 10:40:10 | 8000 | 48 | 52 | 800 | 10:39:09 | S3 | 52 | 800 | 10:39:09 | S3 |
| B4 | 10:40:75 | 3000 | 47 | 52 | 2000 | 10:41:00 | S4 | 52 | 2000 | 10:41:00 | S4 |

**Table 2 Hidden Order Example**

This table highlights the evolution of the order book after the introduction of an iceberg order.

| BUYS | | | SELLS – before | | | | SELLS – after | | | |
|---|---|---|---|---|---|---|---|---|---|---|
| Id | Size | Price | Price | Size | Time | Id | Price | Size | Time | Id |
| | | | | | | | ~~51~~ | ~~2000~~ | ~~10:20:00~~ | ~~S1~~ |
| B1 | 2000 | 50 | 51 | 2000 | 10:20:00 | S1 | ~~51~~ | ~~1000~~ | ~~10:21:15~~ | ~~S2~~ |
| B2 | 700 | 49 | 51 | 2000 | 10:21:15 | S2 | 51 | 1000 | 10:21:15 | S2 |
| B3 | 2500 | 48 | *51* | *18000* | *10:05:10* | *H1* | 51 | 2000 | 10:28:30 | S4 |
| | | | 52 | 2500 | 10:25:31 | S3 | *51* | *16000* | *10:05:10* | *H1* |

**Figure 1**

This figure shows the market impact cost and timing risk as a function of the trading rate. The optimal trading rate is obtained from the minimum point on the curve $\phi(\alpha) + \lambda \mathcal{R}(\alpha)$.

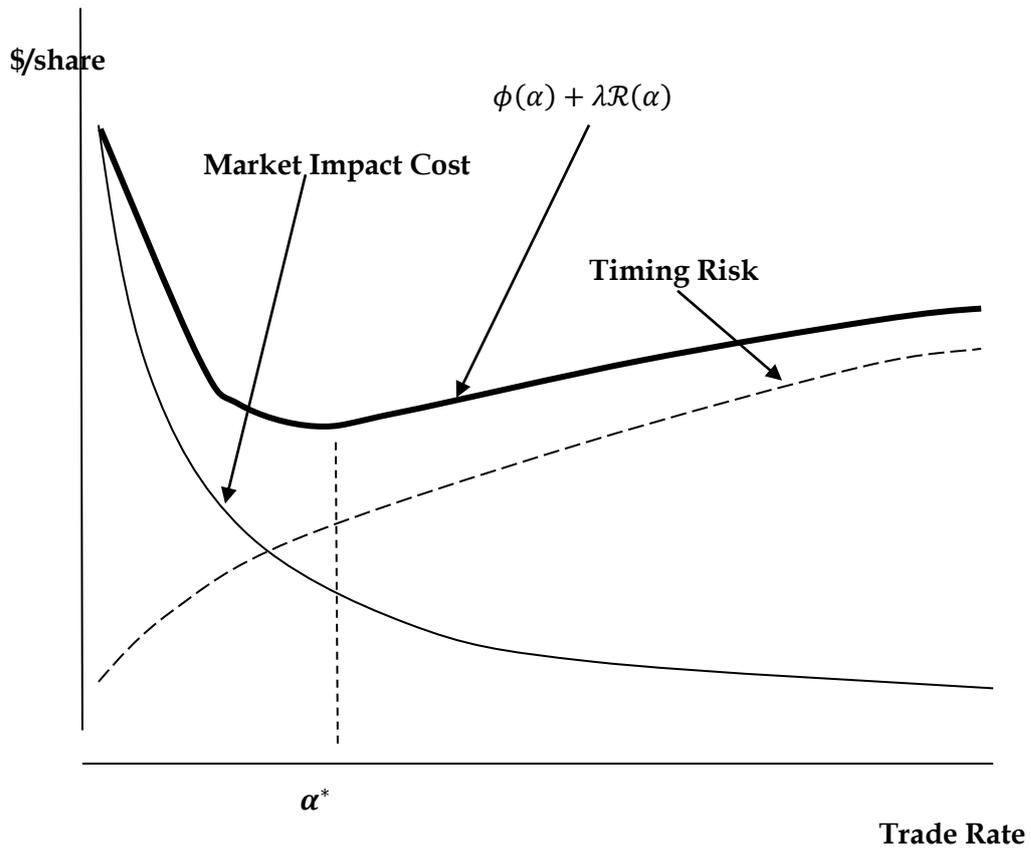

**Figure 2**

This figure shows the ETF computed as conditional to two prices: previous close and arrival price.

**Expected Cost**

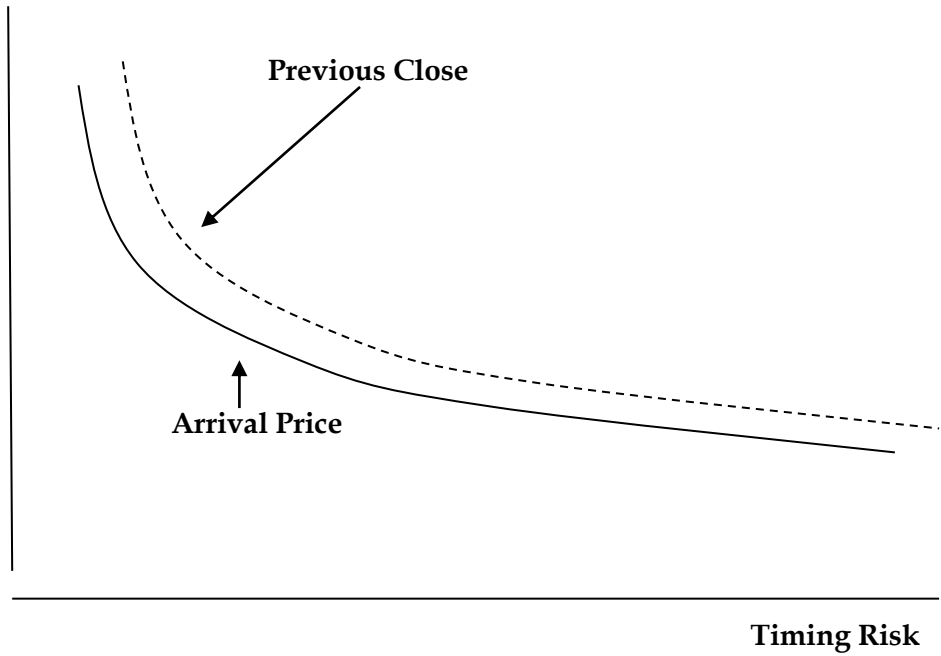

**Timing Risk**

**Table 3 Slicing Execution Tactics Example**

**Panel A. Before: Incoming Market Order to Buy 2200**

| BUYS | | | | SELLS | | | |
|---|---|---|---|---|---|---|---|
| Id | Time | Size | Price | Price | Size | Time | Id |
| B1 | 10:25:00 | 1000 | *50* | *51* | 1000 | 10:20:00 | S1 |
| B2 | 10:20:25 | 2000 | *49* | *51* | 800 | 10:25:25 | S2 |
| B3 | 10:24:20 | 1500 | *48* | *52* | 2500 | 10:24:09 | S3 |

**Panel B. Market Order Crosses**

| SELLS | | |
|---|---|---|
| Price | Size | Id |
| ~~51~~ | ~~1000~~ | ~~S1~~ |
| ~~51~~ | ~~800~~ | ~~S2~~ |
| ~~52~~ | ~~400~~ | ~~S3~~ |
| 52 | 2100 | S3 |

**Panel C. After for a Synthetic Iceberg**

| BUYS | | | | SELLS | | | |
|---|---|---|---|---|---|---|---|
| Id | Time | Size | Price | Price | Size | Time | Id |
| B1 | 10:25:00 | 1000 | *50* | ~~51~~ | ~~1000~~ | ~~10:20:00~~ | ~~S1~~ |
| B2 | 10:20:25 | 2000 | *49* | ~~51~~ | ~~800~~ | ~~10:25:25~~ | ~~S2~~ |
| B3 | 10:24:20 | 1500 | *48* | ~~52~~ | ~~400~~ | ~~10:24:09~~ | ~~S3~~ |
| | | | | 51 | 850 | 10:26:00 | S6 |
| | | | | 52 | 2100 | 10:24:09 | S3 |

**Panel D. For a Native**

| SELLS | | |
|---|---|---|
| Price | Size | Id |
| ~~51~~ | ~~1000~~ | ~~S1~~ |
| ~~51~~ | ~~800~~ | ~~S2~~ |
| ~~51~~ | ~~400~~ | ~~S6~~ |
| 51 | 600 | S6 |
| 52 | 2500 | S3 |

**Table 4 Hiding Execution Tactics Example**

| SELLS | | |
|---|---|---|
| Price | Size | Id |
| *51* | 1000 | S1 |
| *51* | 3000 | S2 |
| **52** | **4000** | **S3** |
| *52* | 2000 | S4 |

| b - SELLS | | |
|---|---|---|
| Price | Size | Id |
| *51* | 2000 | S1 |
| *51* | 3000 | S2 |
| **51** | **1000** | **S3** |
| *51* | 7000 | *H1* |
| *52* | 4000 | S4 |

| c - SELLS | | |
|---|---|---|
| Price | Size | Id |
| *51* | 1000 | S1 |
| *51* | 2000 | S2 |
| *51* | 4000 | *H1* |
| *52* | 1000 | S4 |